\documentclass[usenatbib,usegraphicx]{mn2e}
\usepackage{amsmath}

\title[$\gamma$-ray properties of PKS 1510-089]{Locating the $\gamma$-ray emission region of the flat spectrum radio quasar PKS 1510-089}

\author[Anthony M. Brown]{Anthony M. Brown$^{1}$\thanks{E-mail: anthony.brown@canterbury.ac.nz}\\
$^{1}$Department of Physics and Astronomy, University of Canterbury, Christchurch, 8140, New Zealand}
\begin{document}

\date{Accepted XXX. Received XXX. In original form 2012 November 28}

\pagerange{\pageref{firstpage}--\pageref{lastpage}} \pubyear{2012}

\maketitle

\label{firstpage}

\begin{abstract}
I present a study of the high-energy $\gamma$-ray properties of the flat spectrum radio quasar, PKS 1510-089, based on 3.75 years of observations with the Large Area Telescope (LAT) detector on-board the \textit{Fermi} $\gamma$-ray Space Telescope. Throughout the observing period, the 0.1 GeV $< E_{\gamma} <$ 300 GeV $\gamma$-ray flux was highly variable, undergoing several flaring events where the daily flux exceeded 10$^{-5}$ photons cm$^{-2}$s$^{-1}$ on 3 separate occasions. The increased photon statistics of these large flares allowed the observations to be re-analysed in 6 and 3 hour intervals, revealing flux doubling timescales as small as $1.3\pm0.12$ hours during the flare rise time, and flux halving timescales of $1.21 \pm 0.15$ hours during the flare decay. These are the smallest variability timescales measured to date at MeV$-$GeV energies for the Flat Spectrum Quasar class of Active Galactic Nuclei. 

The $>10^{-5}$ photons cm$^{-2}$s$^{-1}$ flare events were also studied in more detail in an attempt to uncover evidence for the location of PKS 1510-089's $\gamma$-ray emission region. In particular, two approaches were used: (i) searching for an energy dependence to the cooling timescales, and (ii) searching for evidence of a spectral cut-off. The combined results of these two approaches, along with the confirmation of $\geq 20$ GeV photon flux from PKS 1510-089, suggest the presence of multiple $\gamma$-ray emission regions being located in both the broad line region and molecular torus region of PKS 1510-089.

An analysis of the highest photon events within the 3.75 year data set finds PKS 1510-089 to be a source of $\geq20$ GeV $\gamma$-rays at the $13.5\sigma$ confidence level; a observational property which is difficult to explain in the traditional view that $\gamma$-ray emission from Active Galactic Nuclei originates from the base of the relativistic jet. This gives further weight to the argument that there are multiple, simultaneously active $\gamma$-ray emission regions located along the relativistic jet of Active Galactic Nuclei.

\end{abstract}


\section{INTRODUCTION}
The successful launch of the \textit{Fermi} $\gamma$-ray Space Telescope affords us an ideal opportunity to investigate the inner workings of Active Galactic Nuclei (AGN). The ability of the \textit{Fermi}-LAT detector to scan the entire $\gamma$-ray sky every three hours allows us to study the $\gamma$-ray emission from AGN, unbiased by activity state or AGN sub-class. This ability has revealed the blazar subclass of AGN to be the most numerous class of known $\gamma$-ray sources (\cite{nolan}), with approximately equal number of flat spectrum radio quasars (FSRQ) and BL Lac objects (\cite{2fgl}). Furthermore, not only do blazars dominated the extra-galactic $\gamma$-ray sky, but during brief periods of intense flare activity, they can outshine galactic $\gamma$-ray sources as well (\cite{3c454}).

Apart from the radio lobes of the nearby radio galaxy, Centarus A (\cite{cena}), the $\gamma$-ray emission region of AGN remains unresolved. As such, the origin of the $\gamma$-ray emission within blazars remains an open question. Answering this question is an active area of AGN research, with two main schools of thought: on the one hand, some believe the $\gamma$-ray emission originates from within the broad line region (BLR) of the AGN, while on the other hand, some believe the $\gamma$-ray originates further out from the central super-massive black hole (SMBH), within the Molecular Torus (MT) region of the AGN.

Traditionally the $\gamma$-ray emission has been assumed to be close to the base of the relativistic jet, within the BLR. The reasoning for this assumption is two-fold: broad-band spectral energy distribution modelling and rapid $\gamma$-ray flux variability on small timescales.

Multi-wavelength (MWL) observations of FSRQs have found the broad-band spectral energy distribution to be adequately described by a leptonic model, with the emission region located within $300-1000$ Schwarzschild radii from the central SMBH (for example, see \cite{ghietal10} and \cite{nal}). Likewise, the rapid flux variability implies a small emission region size. This is often interpreted as evidence of the emission region being located close to the base of the jet. This interpretation is based on the assumption that the full width of the relativistic jet is responsible for the observed $\gamma$-ray emission and that the size of the emission region, $r$, is simply related to the opening angle of the relativistic jet, $\psi$, and the distance from the SMBH, $R$, via $r \sim \psi R$ (\cite{derm09}, \cite{ghitev09}).

More recently, the wealth of information afford to us by the \textit{Fermi}-LAT detector has found evidence of spectral breaks at GeV energies in the $\gamma$-ray spectrum of some AGN (\cite{specbreak}). These spectral breaks have been interpreted in the context of $\gamma$-ray absorption through photon-photon pair-production with the He Lyman recombination continuum of the BLR (\cite{poutst}), thus pointing to a BLR origin of the observed $\gamma$-ray flux. It is worth noting though, that this interpretation has recently been brought into question (\cite{harris}).

$\gamma$-ray emission from AGN has also been suggested to originate from the MT region of the jet, on the parsec-scale distance from the central SMBH. This suggestion is primarily based on the results of recent MWL observations which have found $\gamma$-ray flaring events to be accompanied by flare events at optical or radio wavelengths, with some of these radio flares being resolved on a parsec-scale distance from the SMBH (for example, see \cite{lah}, \cite{marsch}, \cite{agudo}, \cite{orienti}). 

Further evidence for an MT origin of the $\gamma$-ray emission comes from the detection of Very High Energy (VHE) $\gamma$-ray emission from FSRQs. To date, 3 FSRQs have been detected at energies $\geq 100$ GeV, of which PKS 1510-089 is one (\cite{hess}, \cite{magic2}, \cite{magic3}). The photon-rich environment of the BLR of FSRQs is believed to severely attenuate any $\gamma$-ray emission through photon-photon pair production, resulting in a spectral cut-off above $\geq 20$ GeV (for example, see \cite{donea}, \cite{lui}). As such, the detection of VHE emission is difficult to explain with a pure BLR-origin for the observed $\gamma$-ray.

One of the primary arguments against an MT origin for the $\gamma$-ray emission is that the further an emission region is from the central SMBH, the bigger it is through the combined effects of adiabatic expansion and the opening angle of the jet. However, this argument is only valid if the entire width of the relativistic jet is responsible for the observed $\gamma$-ray emission. What is more, this argument assumes that process of adiabatic expansion is dominant over any re-collimation process that occurs along the length of the jet. The latter assumption is not valid if the jet undergoes re-confinement (e.g., see \cite{soko}). Furthermore, detailed computer simulations have found that jet instabilities can result in large over-densities in the matter distribution at large distances from the central SMBH (for example, see \cite{nish03}, \cite{perc06}, \cite{brom}, \cite{kohler}).

Interestingly, using the above arguments, evidence has been found for the the $\gamma$-ray emission region of PKS 1510-089 to be located in both the BLR and the MT. From the first 11 months of \textit{Fermi}-LAT operation, Abdo et al. concluded that the $\gamma$-ray emission originated from within the BLR (\cite{pks15102010}); this conclusion primarily being driven by the presence of a cut-off in the $\gamma$-ray spectrum. However, both Marscher et al. and Orienti et al. have concluded from their respective MWL campaigns of PKS 1510-089 that the $\gamma$-ray originates from outside BLR region (\cite{marsch}, \cite{orienti}), with their conclusions primarily based on the flaring events being observed simultaneously at $\gamma$-ray, optical and radio wavelengths. It is important to note however, that these three MWL studies focused on three separate flaring events.

This paper investigates the high-energy $\gamma$-ray flux and spectral properties of PKS1510-089 during the first 3.75 years of \textit{Fermi}-LAT observations. In particular, the increased photon statistics associated with several large $\gamma$-ray flare events where the daily flux occasionally exceeds $10^{-5}$ photons cm$^{-2}$s$^{-1}$, allow us to probe flux and spectral variability from PKS 1510-089 with unprecedented temporal resolution in the MeV-GeV energy range. In \textsection 2, I describe the \textit{Fermi}-LAT observations and data analysis routines used in this study. The results on flux variability are shown in \textsection 3 with the in-depth flare analysis reported in \textsection 4. A brief discussion on a multi-zone model is given in \textsection 5 with the conclusions given in \textsection 7. \textsection 6 touches on the VHE $\gamma$-ray properties of PKS 1510-089 in the context of understanding the origin of the observed $\gamma$-ray emission. Throughout this paper, a $\Lambda$ cold dark matter ($\Lambda$CDM) cosmology was adopted, with a Hubble constant of H$_{0}=71$ km s$^{-1}$ Mpc$^{-1}$, $\Omega_{m}=0.27$ and $\Omega_{\Lambda}=0.73$ as derived from \textit{Wilkinson Microwave Anisotropy Probe} results (\cite{cdm}).

\section{\textit{Fermi}-LAT OBSERVATIONS AND DATA REDUCTION}

\begin{figure*}
 \centering
 \begin{minipage}{195mm}
\includegraphics[width=190mm]{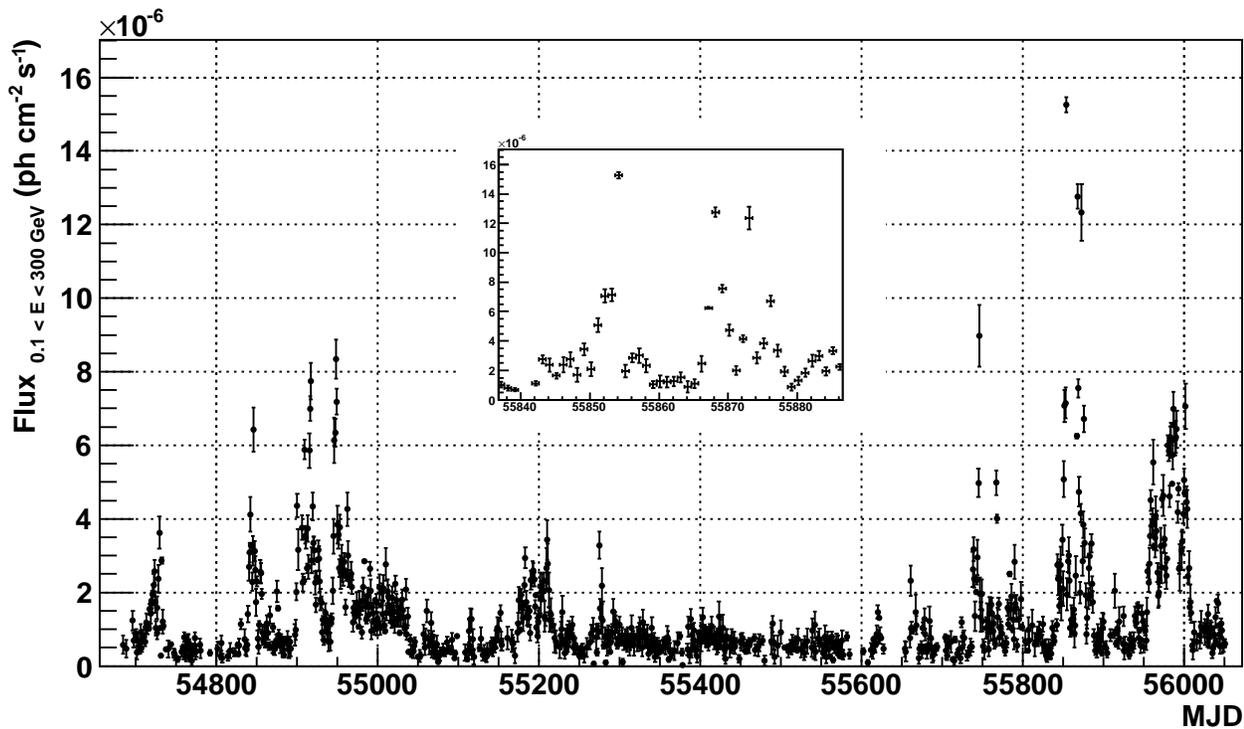}
\caption{Lightcurve of the 0.1 GeV $< E_{\gamma} <$ 300 GeV $\gamma$-ray flux from PKS 1510-089, during the period 2008 August 4 (MJD 54682) to 2012 May 4 (MJD 56051), binned in daily periods. Only daily periods with a $TS\geq10$ are shown. The light curve of the flare period where the flux exceeded $10^{-5}$ photons cm$^{-2}$ s$^{-1}$ is shown in the insert.}
\label{lc1}
\end{minipage}
\end{figure*}

The LAT detector aboard \textit{Fermi}, described in detail by \cite{lat}, is a pair-conversion telescope, sensitive to a photon energy range from below 20 MeV to above 300 GeV. With a large field of view, $ \simeq 2.4 $ sr, improved angular resolution, $\sim0.8$\ensuremath{^{\circ}} at 1 GeV\footnote{Below 10 GeV photon energy, the 68\% containment angle of the photon direction is approximately given by $\theta \simeq 0.8$\ensuremath{^{\circ}}($E_\gamma/$GeV)$^{-0.8}$, with the 95\% containment angle being less than 1.6 times the angle for 68\% containment.}, and large effective area, $\sim 8000$ cm$^{2}$ on axis at 10 GeV, \textit{Fermi}-LAT provides an order of magnitude improvement in performance compared to its \textit{EGRET} predecessor. 

Since 2008 August 4, the vast majority of data taken by \textit{Fermi} has been performed in \textit{all-sky-survey} mode, whereby the \textit{Fermi}-LAT detector points away from the Earth and rocks north and south of its orbital plane. This rocking motion, coupled with \textit{Fermi}-LAT's large effective area, allows \textit{Fermi} to scan the entire $\gamma$-ray sky every two orbits, or approximately every three hours (\cite{ritz}). This observational characteristic of the \textit{Fermi} satellite affords us, for the first time, continuous monitoring of the high-energy $\gamma$-ray sky, allowing us to study the high-energy properties of AGN without suffering the biased of activity state often associated with pointed observations. 

The data utilised in this study comprises of all \textit{all-sky-survey} observations taken during the first 3.75 years of \textit{Fermi}-LAT operation, from 2008 August 4 to 2012 May 4, equating to a mission elapse time (MET) interval of 239557417 to 357818882. In accordance with the \textsc{pass}7 criteria, a zenith cut of 100\ensuremath{^{\circ}} along with a rock-angle cut of 52\ensuremath{^{\circ}} was applied to the data to remove any cosmic ray induced $\gamma$-rays from the limb of the Earth's atmosphere. All `source' class events\footnote{`Source' class events equates to \textsc{event\_class}$=$2 in the \textsc{pass}7 data set.} in a 15\ensuremath{^{\circ}} radius of interest (RoI) centered on PKS 1510-089 were considered in the 0.1 GeV $< E_{\gamma} <$ 300 GeV energy range.

Throughout this analysis, \textit{Fermi}-LAT Science Tools version \textsc{v9r27p1} were used in conjunction with instrument response functions (IRFs) \textsc{p7source\_v6}. To investigate the $\gamma$-ray properties of PKS 1510-089, I utilized the unbinned maximum likelihood estimator of the \textit{Fermi}-LAT Science Tools' \textsc{gtlike} routine. \textsc{gtlike} allows us to fit the data with a series of both point and diffuse sources of $\gamma$-rays. The model used to calculate the likelihood of $\gamma$-ray emission from PKS 1510-089 was a combination of the most recent galactic, gal\_2yearp7v6\_v0.fits, and extragalactic, iso\_p7v6source.txt, diffuse models, and all point sources within a 15\ensuremath{^{\circ}} RoI centered on PKS 1510-089. Each point source was modeled with a power-law spectrum of the form $dN/dE = $ A$ \times (E/E_o)^{-\Gamma}$. Both the photon index, $\Gamma$, and normalisation, A, of point sources within 10\ensuremath{^{\circ}} of PKS 1510-089 were free to vary during the model fitting, while, for sources greater than 10\ensuremath{^{\circ}} from PKS 1510-089, both $\Gamma$ and A were fixed to the values published in the Second Fermi Source Catalog (\cite{nolan}). 

\section{$\gamma$-ray characteristics}

\subsection{Light curve}

To investigate the temporal behavior of the $\gamma$-ray flux, the 3.75 year data set was binned into daily temporal bins, with the \textsc{gtlike} routine applied to each bin separately. Only time intervals where the corresponding test statistic\footnote{The test statistic, TS, is defined as twice the difference between the log-likelihood of two different models, $2[\text{log} L - \text{log} L_{0}]$, where $L$ and $L_{0}$ are defined as the likelihood when the source is included or not respectively (\cite{mattox2}).}, TS, was greater than 10 were considered, which equates to a significance of $\approx 3 \sigma$. The resultant light curve can be seen in Figure 1, with statistical errors only.

Throughout the first 3.75 years of operation, PKS 1510-089 was one of the brightest AGN detected by the \textit{Fermi}-LAT. Due to these high flux levels, requiring $TS \geq 10$ removes few daily bins from Figure 1. The variable nature of PKS 1510-089's $\gamma$-ray flux is clear to see, with the 3.75 year light curve exhibiting short periods of flaring activity separated by extended periods of low fluxes on the order of $(0.5 - 1) \times 10^{-6}$ photons cm$^{-2}$ s$^{-1}$. PKS 1510-089's average daily $\gamma$-ray flux during the observing period is ($1.39 \pm 0.05$) $\times 10^{-6}$ photons cm$^{-2}$ s$^{-1}$, though on 3 separate days, the flux exceeded $10^{-5}$ photons cm$^{-2}$ s$^{-1}$ at the peak of the flare period. During these flares, the $\gamma$-ray flux was at historical maxima for PKS 1510-089.

\begin{figure}
 \centering
\includegraphics[width=95mm]{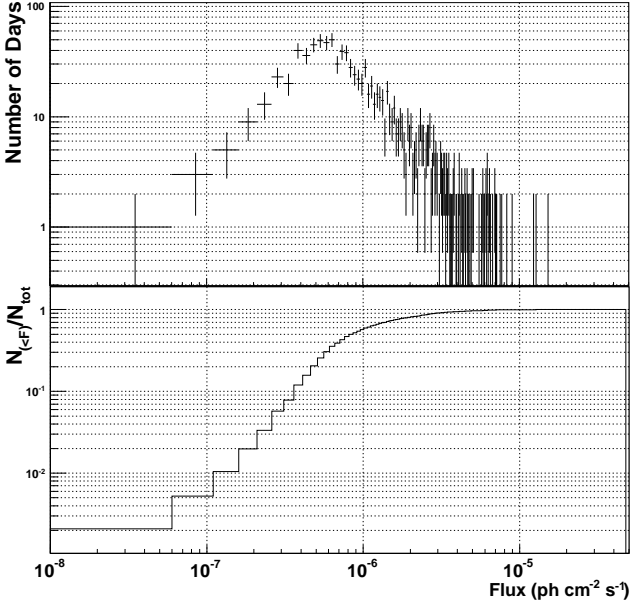}
\caption{\textit{Upper panel}: Duty Cycle of PKS 1510-089's 0.1 GeV $< E_{\gamma} <$ 300 GeV $\gamma$-ray flux during the first 3.75 years of \textit{Fermi}-LAT observations, with the errors simply being the statistical error of the counts in each bin. \textit{Lower panel}: Cumulative distribution function of the duty cycle, showing the number of days where the flux was below a given value F, normalised to the total number of days where the TS value was $TS \geq 10$. The cumulative distribution function clearly shows that the time spent by PKS 1510-089 at extreme flux levels of $> 10^{-5}$ photons cm$^{-2}$ s$^{-1}$ constitutes a small fraction of the total observing period.}
\label{duty}
\end{figure}

To quantify the extent of the variability seen in Figure \ref{lc1}, the duty cycle of the 0.1 GeV $< E_{\gamma} <$ 300 GeV $\gamma$-ray flux during the 3.75 year observing period was derived. Defined as the fraction of time a source is at any given flux level, the duty cycle allows us to investigate if the observed flaring activity is typical or atypical for the source (e.g. \cite{ver04}, \cite{lott12}). Shown in Figure \ref{duty}, the duty cycle utilised all $TS\geq10$ flux measurements from the daily-binned light curve of Figure \ref{lc1}. The duty cycle's bins were defined by the minimum and maximum daily flux values observed, with a total of 100 bins within the flux range.  

The 3.75 year duty cycle of PKS 1510-089 appears to have a well defined peak, where PKS 1510-089 spends the majority of time with a gradual decrease both above and below this peak value, and an excess of events at the highest flux levels associated with the 3 days were the flux was above $10^{-5}$ photons cm$^{-2}$ s$^{-1}$. \cite{tav} constructed a \textit{Fermi}-LAT duty cycle for PKS 1510-089 from 1.5 years of LAT observations, finding that the source was at a high flux level for approximately 1\% of the total observing period, where `high flux level' is defined as a flux that is a factor of 10 larger than the average flux level over the observing period. However, Tavecchio et al. noted that this percentage would most likely decrease over a larger observing period since they studied PKS 1510-089 specifically because it was in a high activity state during the observing period. This does indeed seem to be the case with the exceptional flare events observed during $55850 < $ MJD $<55880$, which are $\sim$10 times the 3.75 year averaged flux, covering only $\sim 0.2$\% of the total observing period reported here. 

\subsection{Flux variability}

\begin{table*}
 \begin{minipage}{150mm}
   \caption{Summary of quickest variability timescales events of PKS 1510-089 during the 3.75 year period, which are less than 3 hours and have a significance of at least $5\sigma$. The times, T$_{start}$ and T$_{stop}$, are in MJD, with the fluxes in units of $10^{-6}$ photons cm$^{-2}$ s$^{-1}$. The observed characteristic timescale $\tau$, from Equation 1, is converted to the intrinsic variability timescale, $\tau_{int}$, with $\tau_{int}=\tau(1+z)^{-1}$. The uncertainty in the variability timescale was calculated by propagating the uncertainty in the flux and time values through Equation 1. The last column indicates whether the variability event is an increase (rise) or decrease (decay) in the flux.}
   \begin{center}
     \begin{tabular}{llllll} \hline \hline
      T$_{start}$ & T$_{stop}$ & Flux start ($F_o$) & Flux stop ($F$) &  $\tau_{int}$ (hours) & Rise/Decay \\ \hline
55850.312 & 55850.437 & $2.97 \pm 1.15$ & $5.00 \pm 1.03$ & $ 2.93 \pm 0.41 $ &  R \\ 
55850.812 & 55850.937 & $2.03 \pm 0.91$ & $4.30 \pm 0.81$ & $ 2.03 \pm 0.39 $ &  R \\ 
55850.937 & 55851.062 & $4.30 \pm 0.81$ & $9.46 \pm 1.07$ & $ 1.94 \pm 0.22 $ &  R \\ 
55851.812 & 55851.937 & $4.75 \pm 1.87$ & $9.48 \pm 1.83$ & $ 2.21 \pm 0.39 $ &  R \\ 
55852.187 & 55852.312 & $13.2 \pm 0.97$ & $3.71 \pm 0.68$ & $ -1.21 \pm 0.15 $ &  D \\ 
55853.062 & 55853.187 & $4.93 \pm 0.45$ & $15.2 \pm 1.12$ & $ 1.36 \pm 0.13 $ &  R \\ 
55853.187 & 55853.312 & $15.2 \pm 1.12$ & $4.84 \pm 0.49$ & $ -1.34 \pm 0.13 $ &  D \\ 
55854.062 & 55854.187 & $9.86 \pm 1.73$ & $5.45 \pm 1.37$ & $ -2.58 \pm 0.32 $ &  D \\ 
55856.062 & 55856.187 & $2.38 \pm 0.19$ & $5.23 \pm 1.32$ & $ 1.94 \pm 0.17 $ &  R \\ 
55856.187 & 55856.312 & $5.23 \pm 1.32$ & $2.79 \pm 1.00$ & $ -2.42 \pm 0.47 $ &  D \\
55866.312 & 55866.437 & $2.92 \pm 0.38$ & $5.41 \pm 2.45$ & $ 2.48 \pm 0.28 $ &  R \\ 
55867.437 & 55867.562 & $7.17 \pm 0.64$ & $3.96 \pm 2.60$ & $ -2.57 \pm 0.20 $ &  D \\ 
55868.812 & 55868.937 & $7.27 \pm 1.98$ & $12.29 \pm 1.32$ & $ 2.91 \pm 0.22 $ &  R \\ 
55869.062 & 55869.287 & $9.62 \pm 0.34$ & $3.27 \pm 0.29$ & $ -1.42 \pm 0.07 $ &  D \\ 
55869.187 & 55869.312 & $3.27 \pm 0.29$ & $8.50 \pm 2.08$ & $ 1.60 \pm 0.19 $ &  R \\ 
55869.687 & 55869.812 & $3.53 \pm 0.80$ & $6.65 \pm 0.34$ & $ 2.42 \pm 0.11 $ &  R \\ 
55870.187 & 55870.312 & $2.95 \pm 1.12$ & $5.82 \pm 0.82$ & $ 2.25 \pm 0.30 $ &  R \\ 
55870.437 & 55870.687 & $8.71 \pm 1.80$ & $3.95 \pm 1.16$ & $ -1.93 \pm 0.38 $ &  D \\ 
55872.062 & 55872.187 & $1.55 \pm 0.47$ & $4.39 \pm 0.43$ & $ 1.47 \pm 0.21 $ &  R \\ 
55872.562 & 55872.687 & $7.28 \pm 1.58$ & $14.17 \pm 2.09$ & $ 2.29 \pm 0.27 $ &  R \\ 
55873.437 & 55873.562 & $8.50 \pm 1.02$ & $4.71 \pm 0.76$ & $ -2.59 \pm 0.22 $ &  D \\ 
55873.687 & 55873.812 & $5.74 \pm 0.61$ & $3.10 \pm 0.58$ & $ -2.48 \pm 0.21 $ &  D \\ 
55874.687 & 55874.812 & $3.51 \pm 1.35$ & $6.67 \pm 0.97$ & $ 2.38 \pm 0.31 $ &  R \\ 
55875.062 & 55875.187 & $2.76 \pm 1.20$ & $4.74 \pm 1.38$ & $ 2.83 \pm 0.54 $ &  R \\ 
55875.437 & 55875.562 & $1.67 \pm 0.85$ & $5.41 \pm 0.29$ & $ 1.30 \pm 0.12 $ &  R \\ 
55875.812 & 55875.937 & $3.97 \pm 0.66$ & $9.87 \pm 1.84$ & $ 1.68 \pm 0.28 $ &  R \\  \hline
    \end{tabular}
  \end{center}
  \label{comparison}
\end{minipage}
\end{table*}

\begin{figure}
 \centering
\includegraphics[width=90mm]{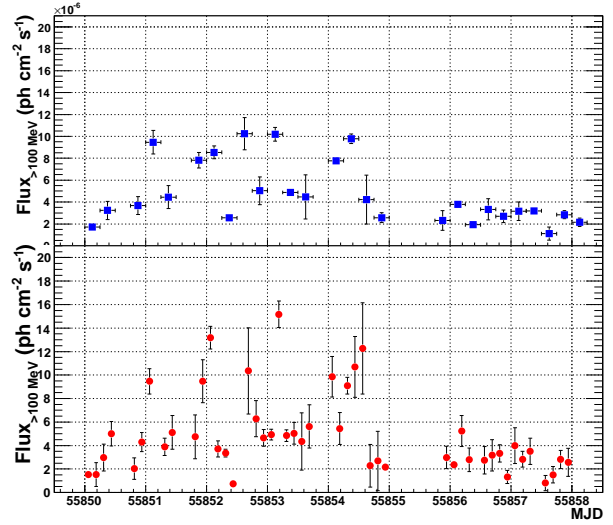}
\caption{\textit{Upper panel}: 6 hour binned light curve of the 0.1 GeV $< E_{\gamma} <$ 300 GeV $\gamma$-ray flux from PKS 1510-089 during the `flare 1' period, $55850 < $ MJD $<55858$. \textit{Lower panel}: 3 hour binned light curve of the 0.1 GeV $< E_{\gamma} <$ 300 GeV $\gamma$-ray flux from PKS 1510-089 during the `flare 1' period, $55850 < $ MJD $<55858$. During this period, the flux exceeds $10^{-5}$ photons cm$^{-2}$ s$^{-1}$ in the 3 hour binned light curve on five separate occasions. All temporal bins have a $TS\geq10$.}
\label{fl1}
\end{figure}

\begin{figure}
 \centering
\includegraphics[width=90mm]{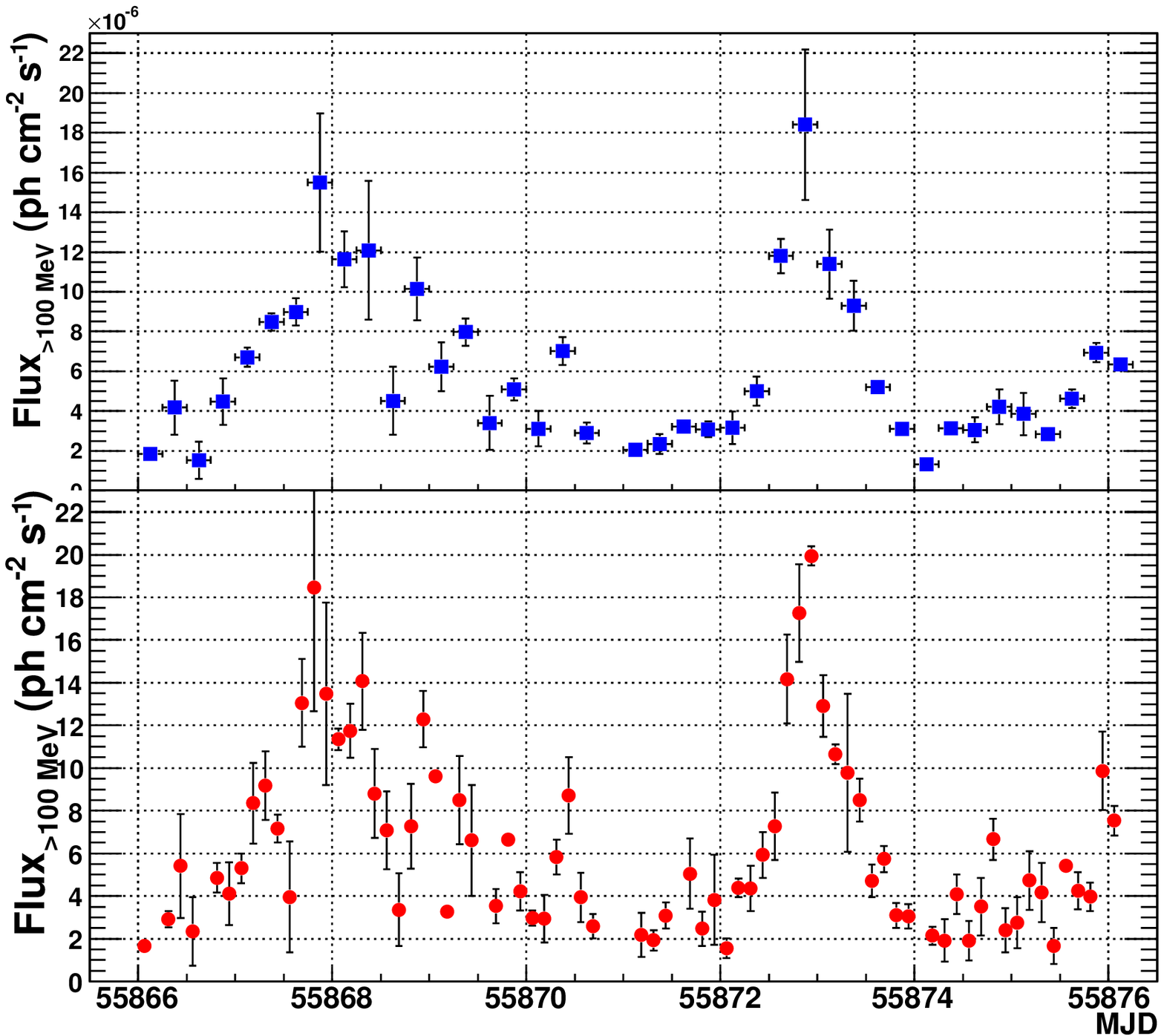}
\caption{\textit{Upper panel}: 6 hour binned light curve of the 0.1 GeV $< E_{\gamma} <$ 300 GeV $\gamma$-ray flux from PKS 1510-089 during the `flare 2' period, $55866 < $ MJD $< 55876$. \textit{Lower panel}: 3 hour binned light curve of the 0.1 GeV $< E_{\gamma} <$ 300 GeV $\gamma$-ray flux from PKS 1510-089 during the `flare 2' period, $55866 < $ MJD $< 55876$. During this period, the flux reached a historic maximum for PKS 1510-089 of $(1.99 \pm 0.04) \times 10^{-5}$ photons cm$^{-2}$ s$^{-1}$ on MJD$=55872.937$. All temporal bins have a $TS\geq10$.}
\label{fl2}
\end{figure}

To search for rapid flux variability on sub-day timescales, the daily light curve shown in Figure \ref{lc1} was utilised to select periods of high flux level. In particular, I concentrated on two flare periods which encompass the three days where the $\gamma$-ray flux from PKS 1510-089 was at historical maximum. The first period, `flare 1', spanning the period $55850 < $ MJD $<55858$, saw the 0.1 GeV $< E_{\gamma} <$ 300 GeV $\gamma$-ray flux from PKS 1510-089 peak at $(1.53 \pm 0.02) \times 10^{-5}$ photons cm$^{-2}$ s$^{-1}$ on MJD $=55854$. The second period, `flare 2', spanned the period $55866 < $ MJD $< 55876$, during which time, PKS 1510-089's daily flux twice exceeded $10^{-5}$ photons cm$^{-2}$ s$^{-1}$; $(1.28 \pm 0.03) \times 10^{-5}$ photons cm$^{-2}$ s$^{-1}$ on MJD $=55868$ and $(1.23 \pm 0.08) \times 10^{-5}$ photons cm$^{-2}$ s$^{-1}$ on MJD $=55873$. These high flux levels allowed us to re-analyse the observations with \textsc{gtlike}, in 6 and 3 hour bins and still satisfy the $TS \geq 10$ criteria for the majority of temporal bins. The resultant light curves can be seen in Figure \ref{fl1} for flare 1 and Figure \ref{fl2} for flare 2. 

As can be seen in both Figure \ref{fl1} and Figure \ref{fl2}, the 3 hour binned light curve reveals a large amount of variability that was masked, or washed out, by the daily bins utilised in Figure \ref{lc1}. Furthermore, the smaller temporal bins of Figure \ref{fl1} and Figure \ref{fl2} reveal that the flux exceeded $10^{-5}$ photons cm$^{-2}$ s$^{-1}$ during 17 separate 3 hour intervals, compared to 3 daily periods revealed in Figure \ref{lc1}, with the 0.1 GeV $< E_{\gamma} <$ 300 GeV $\gamma$-ray flux from PKS 1510-089 peaking at $(1.99 \pm 0.04) \times 10^{-5}$ photons cm$^{-2}$ s$^{-1}$, equating to $\sim 3\times10^{48}$ ergs s$^{-1}$, on MJD$=55872.937\pm0.063$.

To characterise the timescales of the observed flux variability, the time taken for the flux to increase or decrease by a factor of 2 was evaluated. Referred to as the doubling/halving timescale depending upon whether the flux is increasing or decreasing, this timescale is defined by:

\begin{equation}
 F(t)=F(t_{o}) \times 2^{(\tau^{-1}(t-t_{o}))}
\end{equation}

where $\tau$ is the characteristic doubling or halving timescale and $F(t)$ and $F(t_{o})$ are the fluxes at time $t$ and $t_{o}$ respectively. The above equation was applied to consecutive flux measurements that satisfy the $TS \geq 10$ criteria, as shown in Figure \ref{lc1}, Figure \ref{fl1} and Figure \ref{fl2}, in a systematic search for the quickest variability timescale of PKS 1510-089's 0.1 GeV $< E_{\gamma} <$ 300 GeV $\gamma$-ray flux. No selection criteria is applied to the flux level and its associated error; rather, only resultant variability timescales that have a value which is greater than 5 standard deviations from zero were considered. A summary of intrinsic variability timescales with $\mid{\tau_{int}}\mid < 3$ hours, which have a significance of at least 5$\sigma$, are given in Table 1. 

This approach has two important caveats: firstly, for consecutive flux values where the difference is less than a factor of two, the variability timescale calculated is essentially a prediction of how long it would take for a double or halving of flux to occur if the observed change in flux continued at its current rate. Secondly, the variability timescales calculated assumes that the flux increase or decrease is constant throughout the 3 hour bin, and as such, should be considered as an upper limit. Probing timescales smaller than those observed is primarily limited by the 3 hour period that \textit{Fermi}-LAT takes to scan the entire sky and the amount of time a source is in the bore sight of the LAT instrument where the sensitivity is the greatest.

As one would expect, the 3 hour binned light curves of Figure \ref{fl1} \& \ref{fl2} reveal the most rapid variability. In particular, the shortest doubling rise time, $\tau_{rise}$, was found to be $1.30 \pm 0.12$ hours at MJD$=55875.437\pm0.063$, and the shortest halving decay time, $\tau_{decay}$, was $1.21 \pm 0.15$ hours at MJD$=55852.187\pm0.063$. It is worth highlighting that there are several instances of variability on even smaller timescales, for example on MJD$=55852.437\pm0.063$ the intrinsic variability timescale was $-0.91\pm0.66$ hours, however with large uncertainties in the variability timescale, these events do not satisfy the $>5\sigma$ criterion. 

A similar study was undertaken by \cite{fosch} for the FSRQs 3C 454.3, 3C 273 and PKS B$1222+216$ utilising $\approx 2$ years of LAT observations. This study revealed intrinsic flux variability in the $2-3$ hour range, but only at the $3\sigma$ level of certainty. Furthermore, Foschini et al. adopted a conservative approach, and only calculated the variability timescale between subsequent flux measurements with a $3\sigma$ difference in magnitude. If the same approach is applied to the variability events in Table 1, the shortest rise and decay timescales are still found to be $1.30 \pm 0.12$ and $1.21 \pm 0.15$ hours respectively. Indeed, even requiring a $5\sigma$ difference in consecutive flux values to calculate the associated doubling/halving timescale, result in the same timescale for the quickest decay in flux, and only a slightly longer rise time of $1.36 \pm 0.13$ hours. As such, the 1.21 hour decay-time and 1.3 hour rise-time timescales observed in this study are the quickest flux variability observed from the FSRQ subclass of AGN in the MeV-GeV energy range. The implications for the size of this variability is discussed in \textsection 4. 

\section{Origin of the $\gamma$-ray emission}

\subsection{Variability timescale}
Taking the Doppler factor of the relativistic jet into consideration, causality implies that the size of an emission region, R, with a Doppler factor\footnote{$\delta = (\Gamma(1-\beta\text{cos}))^{-1}$ where $\Gamma$ is the bulk Lorentz factor of the jet, $\beta=v/c$ and $\theta$ is the angle to the line of sight.} $\delta$, is related to the $\gamma$-ray variability timescale, $t_{var}$, by:

\begin{equation}
 R \leq ct_{var}\delta(1+z)^{-1}
\end{equation}

where $z$ is the redshift of the source. For the intrinsic variability timescales outlined in Table 1, the size of the $\gamma$-ray emission region for the observed flares can be constrained to be $R\delta^{-1} \leq (0.9-2.3)\times 10^{14}$ cm. For comparison, adopting a SMBH mass of $5.4 \times 10^8 \text{ M}_{\sun}$ from \cite{pks15102010}, the Schwarzschild radius of PKS 1510-089 is $\sim 1.6 \times 10^{14}$ cm.

Radio observations of PKS 1510-089 during February to June 2010 have shown that structures within its relativistic jet have Doppler factors of $\delta= 47$ (\cite{Kadota}), while observations during 1998 to 2001 show similiar values of $\delta = 34-42$ (\cite{jorstad}). If one assumes these rather extreme values for the Doppler factor are a good representation of the average velocity structure within PKS 1510-089's relativistic jet, the emission regions of the flares reported here are found to be less than $3\times10^{-3}$ parsecs in size. This suggests that only a very small portion of the relativistic jet is responsible for the observed $\gamma$-ray flares. 

While these rapid variability timescales allow us to gain insights into the physical properties of the $\gamma$-ray emission region, the extreme nature of the variability does little to constrain the location of the emission region. Small structure within the relativistic jet is not unique to just the inner regions of the AGN; small over-densities of plasma can occur in the jet at both the sub-parsec, BLR, and the parsec-scale, MT region (\cite{nish03}, \cite{perc06}, \cite{brom}, \cite{kohler}). 

External to the jet however, the environment of the BLR and MT are significantly different, with the energy distribution of the BLR photon field peaking at UV energies, while in the MT it peaks at IR energies. This difference in photon energies, and the subsequent cooling effect they have on the energy distribution of the particle population within the emission region, can be utilised to constrain the origin of the $\gamma$-rays.

\cite{tav} exploited this difference to constrain the cooling timescales for a relativistic electron population located within the BLR and within the MT region. Utilising Equation 3 below, Tavecchio et al. calculated the cooling timescale for an electron producing a 100 MeV $\gamma$-ray photon, through the Inverse-Compton (IC) process, to be $t_{cool}^{obs} \approx 800$ seconds in the BLR and $t_{cool}^{obs} \approx 12,000$ seconds in the MT. Furthermore, for a 5 keV X-ray photon produced via the IC process, Tavecchio et al. calculated an electron cooling timescale of $t_{cool}^{obs} \approx 31$ hours in the BLR and $t_{cool}^{obs} \approx 20$ days in the MT\footnote{Given that X-ray cooling timescales from blazars have been observed on the sub-hour timescale, (for example, see \cite{fosch06} for FSRQs and \cite{mephd} for BL Lac objects), the application of this approach to observed X-ray flux variability is questionable.}.

\begin{equation}
 t_{cool}^{obs} = \frac{3m_ec(1+z)}{4\sigma_TU^{\prime}\gamma\delta}
\end{equation}

The $3-6$ hour cooling timescales that Tavecchio et al. observed did not allow them to distinguish between a BLR or a MT origin for the emission region. However utilising the approach outlined in \cite{tav}, the $\tau_{decay} = 1.21 \pm 0.15$ hour cooling timescale discovered in this study would suggest that the emission region, for the quickest flaring event, is located within the BLR\footnote{It should be noted that this conclusion does not apply to all variability timescales reported in Table 1 since, for events with $\tau_{decay} \geq 2.5$ hours, it is not possible to differentiate between the BLR and MT cooling timescales of 800 and 12,000 seconds respectively.}. However, it is important to highlight that this approach assumes that the IC interaction is occuring in the Thomson regime, which as discussed in \textsection 4.2, is mostly likely not the case. As such, this approach might not give a realistic cooling timescale, which would bring the conclusion about the location of the emission region into question.

\begin{figure*}
 \centering
 \begin{minipage}{190mm}
\includegraphics[width=190mm]{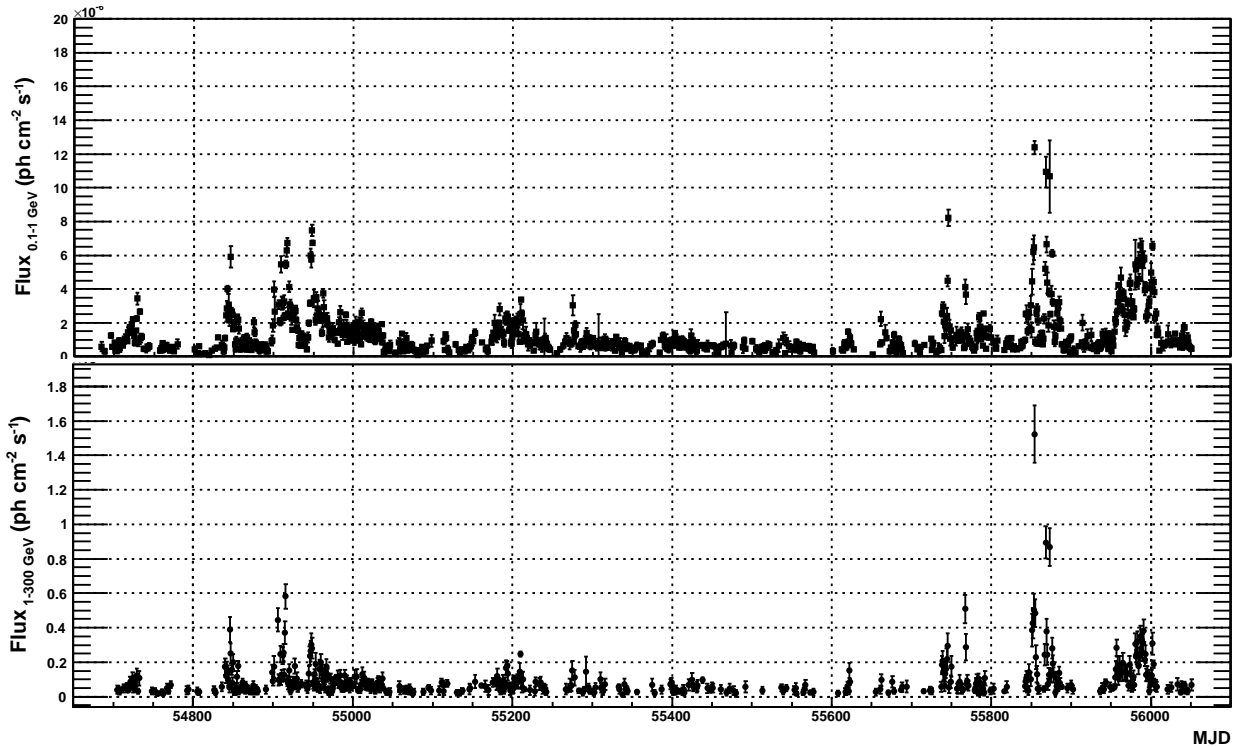}
\caption{\textit{Upper panel}: Lightcurve of the 0.1 GeV $< E_{\gamma} <$ 1 GeV $\gamma$-ray flux from PKS 1510-089, during the period 2008 August 4 (MJD 54682) to 2012 May 4 (MJD 56051), binned in daily periods. \textit{Lower panel:} Lightcurve of the 1 GeV $< E_{\gamma} < $300 GeV $\gamma$-ray flux from PKS 1510-089, during the period 2008 August 4 (MJD 54682) to 2012 May 4 (MJD 56051), binned in daily periods. All temporal bins have a $TS\geq10$.}
\label{lc2}
\end{minipage}
\end{figure*}

\begin{figure}
 \centering
\includegraphics[width=90mm]{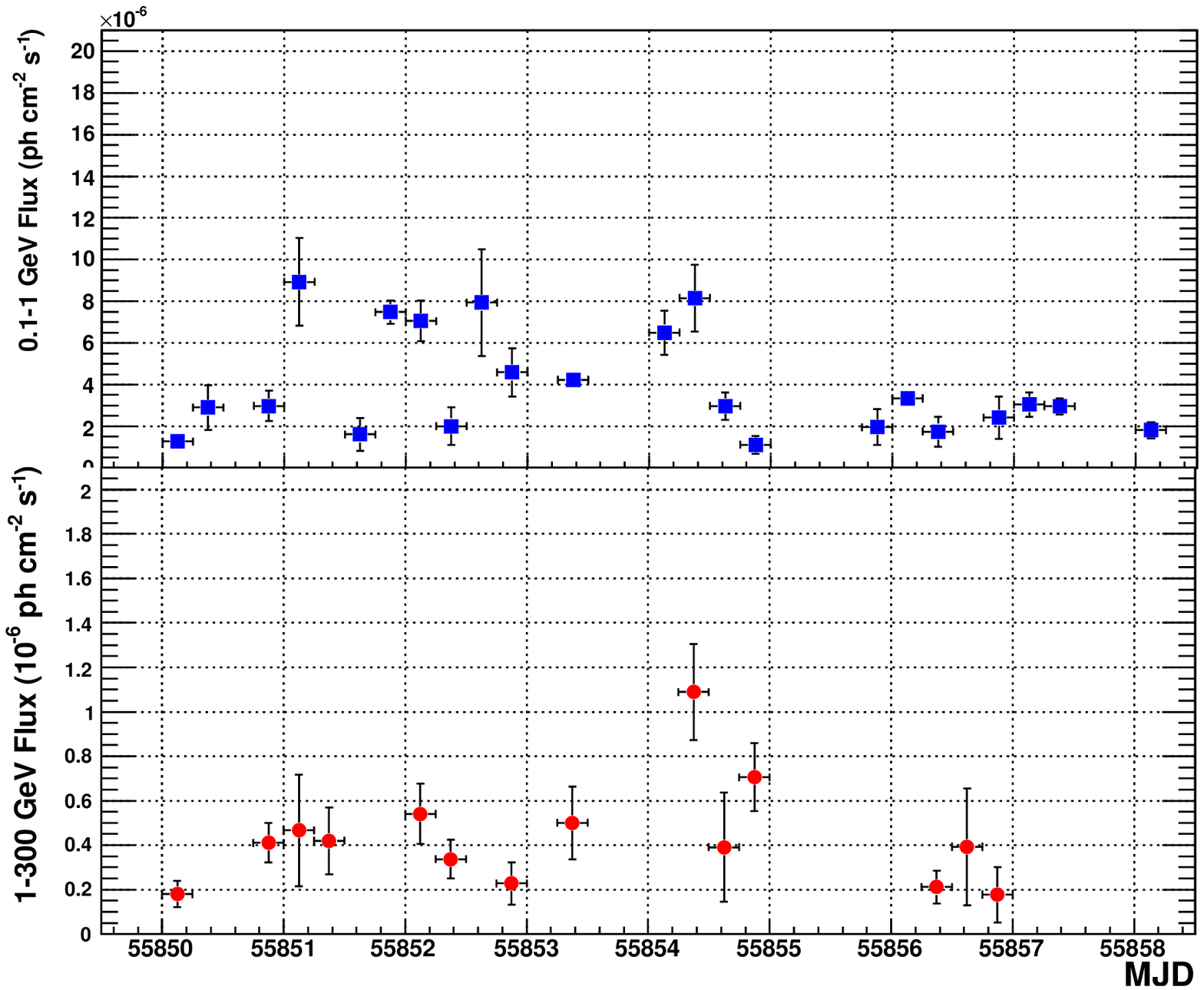}
\caption{\textit{Upper panel}: the 0.1 GeV $< E_{\gamma} <$ 1 GeV $\gamma$-ray flux from PKS 1510-089 during the `flare 1' period, $55850 < $ MJD $<55858$, binned in 6 hour periods. \textit{Lower panel}: the 1 GeV $< E_{\gamma} < $300 GeV $\gamma$-ray flux from PKS 1510-089 during the `flare 1' period, $55850 < $ MJD $<55858$, binned in 6 hour periods. All temporal bins have a $TS\geq10$.}
\label{2efl1}
\end{figure}

\begin{figure}
 \centering
\includegraphics[width=90mm]{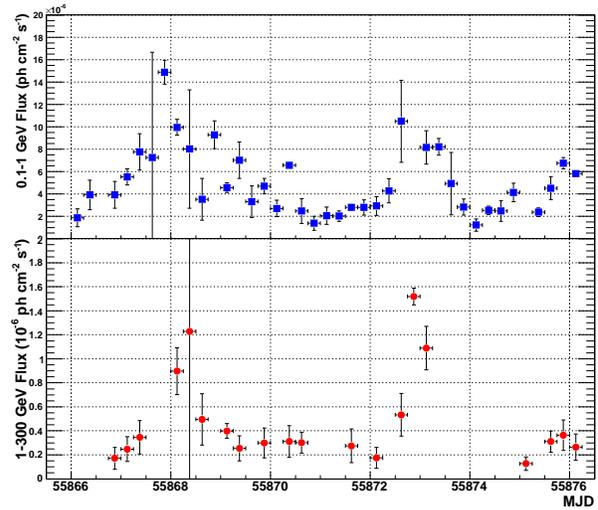}
\caption{\textit{Upper panel}: the 0.1 GeV $< E_{\gamma} <$ 1 GeV $\gamma$-ray flux from PKS 1510-089 during the `flare 2' period, $55866 < $ MJD $< 55876$, binned in 6 hour periods. \textit{Lower panel}: the 1 GeV $< E_{\gamma} < $300 GeV $\gamma$-ray flux from PKS 1510-089 during the `flare 2' period, $55866 < $ MJD $< 55876$ binned in 6 hour periods. All temporal bins have a $TS\geq10$.}
\label{2efl2}
\end{figure}

\subsection{Energy dependent cooling timescales}
Tavecchio et al. assumed that the IC scattering was occuring in the Thomson regime, with $\gamma\epsilon << 1$ being true for all electron-photon interactions, where $\gamma$ is the Lorentz factor for the electron and $\epsilon$ is the energy of the photon. Dotson et al. (2012a,b) on the other hand, found that the IC scattering occurred in the Klein-Nishina regime when the emission region was located in the BLR, while it occurred in the Thomson regime for emission regions located within the MT region of the AGN. This difference manifests itself as an energy dependent cooling timescale for emission regions embedded in the MT region and a (quasi-) energy independent cooling timescale for emission regions embedded in the BLR. Specifically, utilising the BLR and MT photon energy densities from \cite{ghitev09}, Dobson et al. calculated a $<1$ hour difference in the flux halving timescale for 200 MeV photons and 20 GeV photons for an emission region located in the BLR, but a $\sim 10$ hour difference in flux halving timescales for the two photon energies from an emission region embedded in the MT region (\cite{dobson}). 

More generally speaking, this difference would result in a time lag between the cooling of the MeV and GeV components of a $\gamma$-ray flare. Conversely, if present, a time lag can be used to constrain the energy density of the MT photon field (\cite{dotson2}).

To investigate the possibility of energy dependent cooling timescales, the daily light curve of Figure 1, along with the 6 hour binned light curves of Figure 3 and Figure 4, were re-analysed as per the procedure outlined in \textsection 2, but for photons in two distinct energy groups; low-energy, $0.1-1$ GeV, and high-energy, $1-300$ GeV. Due to the limited number of events above 1 GeV and the requirement that each temporal bin has $TS\geq10$, this re-analysis was not applied to the 3 hour binned light curve. The resultant daily binned light curves for both low and high energy fluxes, can be seen in Figure \ref{lc2}. The low and high energy light curve, for both flare 1 and flare 2, binned in 6 hour intervals, are shown in Figure \ref{2efl1} and Figure \ref{2efl2} respectively. 

Discrete correlation functions, (DCF; \cite{ede}), were applied to the low and high energy light curves of Figure \ref{2efl1} and Figure \ref{2efl2} to search for the presence of a time lag in the flux level changes between the energy bands. An important characteristic of the DCF procedure is that it allows us to accommodate for differences in the sampling rates of the two light curves associated with the $TS \geq 10$ criterion. The DCFs represent a more robust approach to searching for energy (in)dependent cooling timescales than simply applying Equation 1 to the individual light curves since the latter is limited by the statistics and associated error of the high energy light curve. Indeed, applying Equation 1 to the daily binned light curves of Figure \ref{lc2} does not find any $1-300$ GeV flux cooling timescales at a $5\sigma$ level of confidence. Furthermore, if the application of Equation 1 reveals the presence of significant variability timescales in both the $0.1-1$ and $1-300$ GeV light curves, \textit{a-priori} knowledge of the magnitude of the time lag is needed to determine if the variability timescales from the individual energy ranges are related.

\begin{figure}
 \centering
\includegraphics[width=100mm]{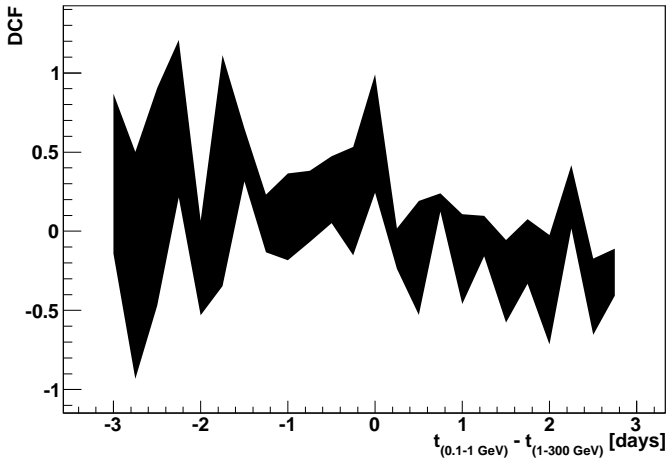}
\caption{Discrete correlation function calculated between the $0.1-1$ GeV and the $1-300$ GeV light curves of Figure \ref{2efl1} for the flare 1 period, binned in 6 hour intervals. The shading indicates the error of the DCF. The DCF indicates that no time lag is present between the two light curves during this flare period.}
\label{dcf1}
\end{figure}

\begin{figure}
 \centering
\includegraphics[width=100mm]{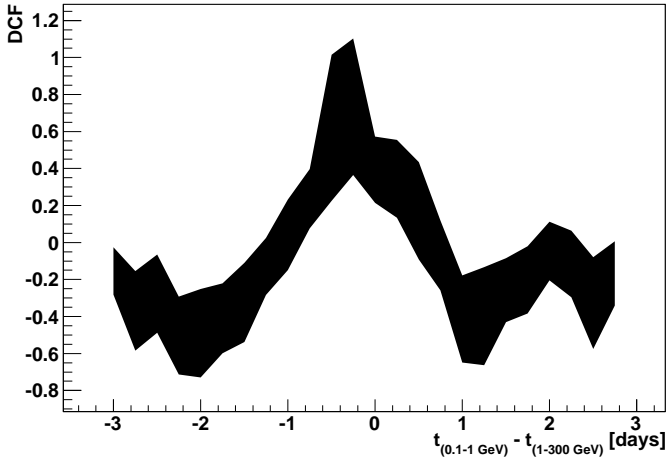}
\caption{Discrete correlation function, calculated between the $0.1-1$ GeV and the $1-300$ GeV light curves of Figure \ref{2efl2} for the flare 2 period, binned in 6 hour intervals. The shading indicates the error of the DCF. A positive time lag implies that the $0.1-1$ GeV flux is delayed with respect to the $1-300$ GeV flux. A gaussian fit to the DCF peak finds a time lag of $-6.13$ hours to be present.}
\label{dcf2}
\end{figure}

The resultant DCFs for the flare 1 and flare 2 events are shown in Figure \ref{dcf1} and Figure \ref{dcf2} respectively, binned in 6 hour intervals. A positive time lag, $t_{(0.1-1 GeV)} - t_{(1-300 GeV)}$, between the two light curves implies that the $0.1-1$ GeV flux is delayed with respect to the $1-300$ GeV flux. However, it is important to note that this delay applies to both the flux increases and flux decreases. To interpret any observed lag in the DCF as evidence for energy dependent cooling timescales, one has to assume that there the flux increase in both energy bands occurs at the same time. 

The DCF for the flare 2 event exhibits a clear peak. A gaussian fit to this peak indicates that this temporal lag is $ -6.13 \pm 2.83$ hours\footnote{It is worth noting that this peak feature is still present if the DCF is rebinned into 12 hour intervals, while the 12 hour binned DCF for flare 1 still has no well defined peak, and large errors for negative time lags.}. However the location of the peak suggests that changes in the $0.1-1$ GeV flux in fact \textbf{proceed} any changes in the $1-300$ GeV flux. A closer inspection of the light curves in Figure \ref{2efl2} finds that the $0.1-1$ GeV flux peaks approximately $6-12$ hours before the $1-300$ GeV flux. As such, it would appear that, if an energy dependent cooling timescale is present, it is masked by the energy dependent rise time of the flare event.

The DCF for flare 1 on the other hand, appears to have a peak at a time-lag of 0. This would suggest that there is no energy dependence to the cooling timescale, and therefore point towards a BLR-origin for the $\gamma$-ray emission. However there is a large amount of uncertainty in the DCF for negative time-lags, with additional peaks at $\sim-2$ and $\sim-2.5$, and thus it is not possible to draw strong conclusions about the DCF peak at 0 and the DCF for flare 1 in general. Furthermore, the DCF for the flare 2 event suggests that the DCF could be dominated by effects not related to the electron cooling timescales. As such, this implies that for these flare events, little is gained from the DCF approach outlined here with regards to determine the location of the $\gamma$-ray emission region. 

\subsection{Photon-photon pair-production}
Another important difference between the two possible locations for the $\gamma$-ray emission regions is the role that photon-photon pair-production, ($\gamma\gamma\rightarrow e^+e^-$) plays in attenuating the emitted $\gamma$-ray flux and the subsequently imprint that the process leaves on the observed $\gamma$-ray spectrum (\cite{donea}, \cite{lui}). Simply put, as a result of photon-photon pair-production, the BLR of FSRQs is opaque to $\gamma$-rays above $\sim20$ GeV in energy, while the MT region is not. As a result, if the $\gamma$-ray region is located in the BLR, one would expect to see a cut-off in the $\gamma$-ray spectrum attributed to pair production, while a $\gamma$-ray spectrum originating from the MT would not have such a feature.

To search for a spectral cut-off feature, the flare 1 and flare 2 light curves were again re-analysed in daily bins, with PKS 1510-089 modeled by 3 additional models to the power-law used originally in \textsection 2. To improve statistics, especially at the high-energy tail of the spectrum, the observations were re-analysed in daily intervals. The additional models utilised were a broken power-law, a broken power-law with an exponential cut-off and a log-parabola. All other point and diffuse sources in the RoI were modeled as before (see \textsection 2). The TS value for each model, in each daily bin, was compared to that of a simply power-law, to investigate if there is a significant deviation from a power-law spectrum during the flaring event. The difference in TS is defined as TS$_x -$ TS$_{power-law}$, where $x$ is the TS value for either the broken power-law, broken power-law with an exponential cut-off or log-parabola fits. A large positive difference in TS values indicates that the $\gamma$-ray spectrum is better described by a function with a cut-off feature then a simple power-law distribution, and as such, suggests the presence of a spectral cut-off. An important caveat of this approach is that the presence of a spectral cut-off does not automatically imply a BLR origin for the $\gamma$-ray flux; for example a cut-off in the $\gamma$-ray spectrum can also occur if there is a cut-off in the energy distribution of the particle population responsible for the $\gamma$-ray emission.

\begin{figure}
 \centering
\includegraphics[width=80mm]{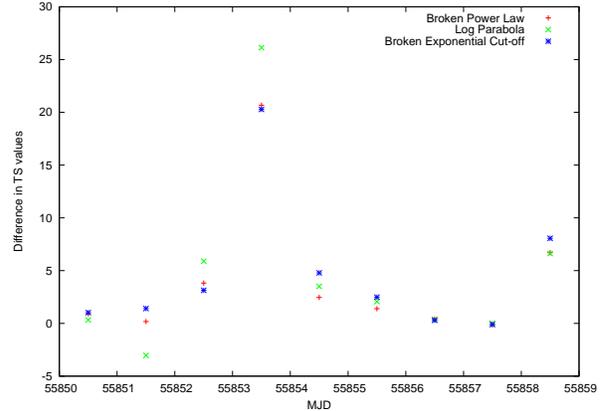}
 \caption{The difference in TS values, defined as TS$_x -$ TS$_{power-law}$, where $x$ is the TS value for either the broken power-law, a broken power-law with an exponential cut-off and a log-parabola fits, throughout the `flare 1' period. To improve statistics, especially at the high-energy tail of the spectrum, the difference in model TS values was calculated on daily intervals of observations. There is a $TS>25$ difference between a power-law and a log-parabola description of PKS 1510-089 during MJD$=55853$.}
\label{ts1}
\end{figure}

\begin{figure}
 \centering
\includegraphics[width=80mm]{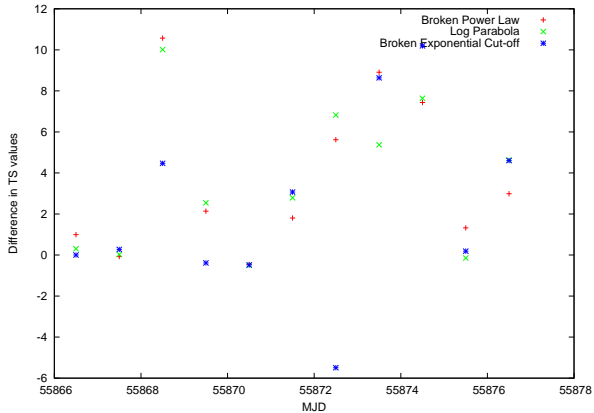}
\caption{The difference in TS values, defined as TS$_x -$ TS$_{power-law}$, where $x$ is the TS value for either the broken power-law, a broken power-law with an exponential cut-off and a log-parabola fits, throughout the `flare 2' period. To improve statistics, especially at the high-energy tail of the spectrum, the difference in model TS values was calculated on daily intervals of observations. No significant difference in TS value is seen throughout the second flare.}
\label{ts2}
\end{figure}

The daily differences in TS values during the flare 1 and flare 2 events can be seen in Figure \ref{ts1} and Figure \ref{ts2} respectively. Flare 1 exhibits a large TS difference, $\Delta TS>25$, between a power-law and a log-parabola description of PKS 1510-089's 0.1 GeV $< E <$ 300 GeV $\gamma$-ray spectrum during MJD$=55853$, a day before flare 1's maximum flux. However, on MJD$=55850$, at the start of flare 1, there is little difference in the TS value from either a power-law or log-parabola description of PKS 1510-089's spectrum. Likewise, towards the end of the flare 1 event, there is also little difference in the TS values of the different spectral shapes. This variation in the TS value for the log-parabola description of the spectrum, relative to the power-law fit, suggests that throughout the flare 1 event, there is a change in the shape of the $\gamma$-ray spectrum. What is more, the preference of the log-parabola fit over the power-law fit implies the presence of a cut-off in the spectrum. 

\begin{figure}
 \centering
\includegraphics[width=90mm]{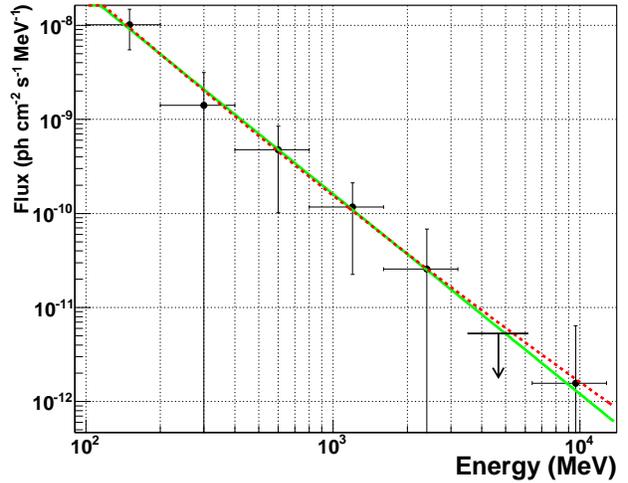}
\caption{The spectrum of PKS 1510-089 during MJD$=55850$. The solid green line indicates the best power-law fit to the spectrum, while the dash red line indicates the best log-parabola fit to the spectrum. The reduced $\chi^2$ of the power-law fit is 0.112 while the reduced $\chi^2$ of the log-parabola fit is 0.155. The \textsc{gtlike} fit to the $3.2-6.4$ GeV energy band has a TS value less than 10, and as such, was replaced by a 95 \% confidence upper limit, however this limit was not included in the reduced $\chi^2$ calculation for the function fit.}
\label{sp1}
\end{figure}

\begin{figure}
 \centering
\includegraphics[width=90mm]{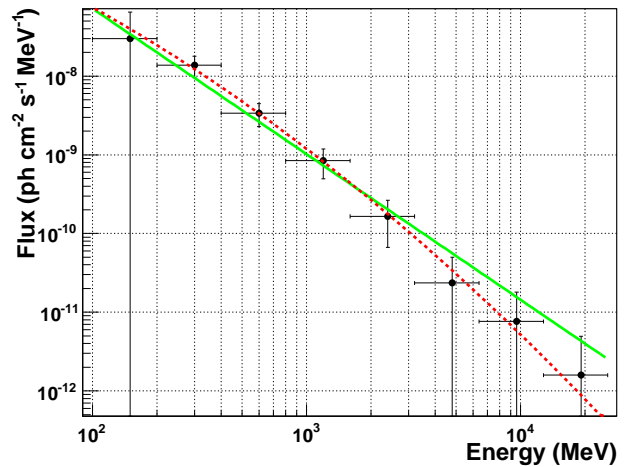}
\caption{The spectrum of PKS 1510-089 during MJD$=55853$. The solid green line indicates the best power-law fit to the spectrum, while the dash red line indicates the best log-parabola fit to the spectrum. The reduced $\chi^2$ of the power-law fit is 1.885 while the reduced $\chi^2$ of the log-parabola fit is 0.282.}
\label{sp2}
\end{figure}

To investigate this spectral cut-off, the $\gamma$-ray spectrum during MJD$=55850$ and MJD$=55853$ were fitted with both a power-law and log-parabola function, and the reduced $\chi^2$ of the function fit was calculated. The spectra were obtained by applying \textsc{gtlike} separately to 10 logarithmic energy bins in the 100 MeV to 102.4 GeV energy range. Only energy bins with a \textsc{gtlike} TS value greater than 10 were considered in the function fit. The resultant spectrum, power-law and log-parabola fits for MJD$=55850$ and MJD$=55853$ can be seen in Figure \ref{sp1} and Figure \ref{sp2} respectively.

From Figure \ref{sp1}, it can be seen that at the start of the flare 1 event, when there is no difference in the TS values, the spectrum is well described by a power-law, with no indication of a spectral cut-off present\footnote{It is important to note that the log-parabola function, which is defined by $\frac{dN}{dE} = A\times E^{(\alpha - \beta \times log10(E))}$, also describes the spectrum well, simply because the co-efficient $\beta$ is close to zero, thus rendering the log-parabola function, a power-law function.}. However the $\gamma$-ray spectrum during MJD$=55853$, seen in Figure \ref{sp2}, exhibits a clear deviation from a simple power-law distribution at the high-energy tail of the spectrum. The deficiency of the power-law is confirmed by the reduced $\chi^2$ of the power-law fit, which is 1.885, while the reduced $\chi^2$ of the log-parabola fit is 0.282. The reduced $\chi^2$ of the log-parabola fit thus confirms the presence of a cut-off in the $\gamma$-ray spectrum for the flare 1 event, and therefore suggests that the flare 1 event originates from within the BLR. 

For flare 2, on the other hand, Figure \ref{ts2} indicates that there is no significant deviation from a power-law description throughout the flare event, with the greatest difference in TS values being $\sim10$. What is more, for a large number of days, the TS value for the power-law fit is greater than that of the spectral models with cut-offs. As such, the lack of a spectral cut-off suggests an MT origin for the $\gamma$-ray emission. 

\section{Multiple emission regions}
Taken in isolation, the presence or lack of a spectral cut-off or temporal lag in the discrete correlation functions, do not provide any strong evidence as to the location of the emission region. However, the combination of the two properties allows for a stronger argument as to the possible location.  

\textbf{Flare 1:} At its peak flux, the $0.1-300$ GeV spectrum of flare 1 ($55850 <$ MJD $< 55858$) showed significant deviation, greater than $TS>25$, from a power-law description, indicating the presence of a spectral cut-off. The presence of this cut-off was confirmed by the reduced $\chi^2$ values for the power-law and log-parabola fits to the $0.1-300$ GeV spectrum on MJD$=55853$, with the power-law fit having a reduced $\chi^2$=1.885 and the log-parabola fits having a reduced $\chi^2$=0.282. Due to photon-photon pair-production, the spectral cut-off suggests that the $\gamma$-ray emission region associated with this flare is located within the BLR, on the sub-parsec scale from the central SMBH. While a peak at a time-lag of 0 in the $0.1-1$ GeV versus $1-300$ GeV DCF for flare 1 would suggest that there is no energy dependence to the cooling timescale, and thus a BLR-origin for the $\gamma$-ray emission, the uncertainty in the DCF for negative time-lags is too large to draw strong conclusions. Furthermore, the DCF for the flare 2 event suggests that the DCF is dominated by effects not related to the electron cooling timescales and as such, implies that little is gained from the DCF approach outlined here with regards to determining the location of the $\gamma$-ray emission region.

\textbf{Flare 2:} During the the flare 2 period ($55866 <$ MJD $< 55876$) the daily flux twice exceeded $10^{-5}$ photons cm$^{-2}$ s$^{-1}$. However there was no significant deviation from a power-law distribution at any point during the flare event. As such, without invoking the presence of axion-like particles\footnote{If the flare 2 event was in fact located within the BLR, axion-like particles could possibly explain the absence of a spectral cut-off (e.g. see \cite{tavax}, \cite{horns}, \cite{ron}). Without axion-like particles, the combination of BLR origin and lack of a spectral cut-off would require a sudden drop in the local energy density of the BLR photon field.}, the lack of spectral cut-off points towards a MT location for the $\gamma$-ray emission region associated with this flare. The $0.1-1$ GeV versus $1-300$ GeV DCF for flare 2 does not reveal any evidence for an energy dependence to the cooling timescales, though this may to be due to the fact that the DCF is dominated by an apparent energy dependence in the flare rise time. 

While there is $\sim 8$ days between the two flaring episodes, there is evidence indicating that the two flare events are spatially separated by $\geq1$ parsec, with flare 1 being of BLR origin and flare 2 being of MT origin. Thus a natural conclusion to draw from these studies is that there are multiple, simultaneously active, $\gamma$-ray emission zones along the relativistic jet in both the BLR and MT, capable of emitting $0.1-300$ GeV $\gamma$-rays. 

Multi-zone emission models have often been used to explain the broad-band spectral energy distribution of blazars (for example, see \cite{mephd}, \cite{nal}), or to explain the $\gamma$-ray emission from misaligned AGN (\cite{lenain}). However, for these multi-zone models, the $\gamma$-ray emission origin is either in the BLR or the MT, never both.  

From a multi-wavelength (MWL) campaign of PKS 1510-089, during the first 11 months of \textit{Fermi}-LAT operation, Abdo et al. concluded that the $\gamma$-ray emission originated from within the BLR (\cite{pks15102010}); this conclusion primarily being driven by the presence of a cut-off in the $\gamma$-ray spectrum. However, both Marscher et al. and Orienti et al. have concluded from their respective MWL campaigns of PKS 1510-089 that the $\gamma$-ray originates from outside BLR region (\cite{marsch}, \cite{orienti}), with their conclusions primarily based on the flaring events being observed simultaneously at $\gamma$-ray, optical and radio wavelengths. The presence of multiple $\gamma$-ray emission regions along the jet allows us to reconcile these seemingly contradictory conclusions, with PKS 1510-089 being able to produce $\gamma$-ray flares from regions both within the BLR and the MT regions of it's relativistic jet.

\section{VHE emission}

\begin{figure}
 \centering
\includegraphics[width=70mm,angle=270]{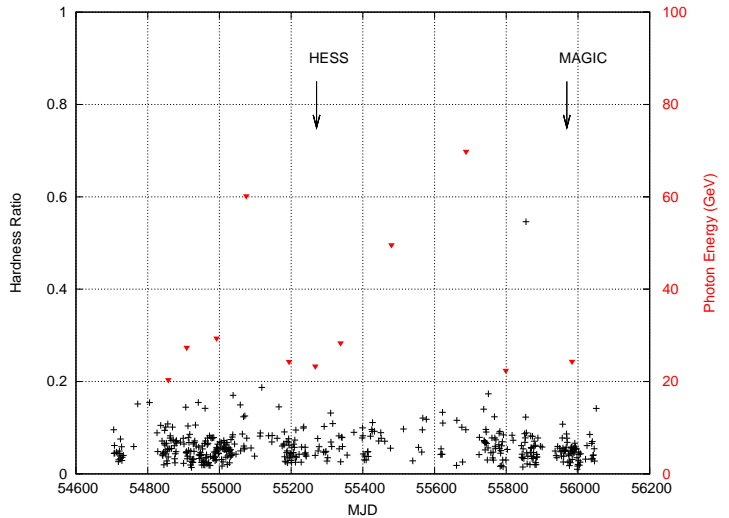}
\caption{The black crosses represent hardness ratio, defined as the $1-300$ GeV flux divided by the $0.1-1 GeV$ flux, of PKS 1510-089 over the entire 3.75 year period. The red filled triangles represent the arrival times of all photons $> 20$ GeV in energy within 0.1 degrees of PKS 150-089. The scale for the hardness ratio is on the left, while the photon energy scale is on the right. There does not appear to be any preference for the highest energy photons to arrive during hard spectral states. For comparison, the MJD when VHE emission was detected by the ground-based HESS and MAGIC telescopes are indicated by the vertical arrows.}
\label{hardphotons}
\end{figure}

Finally I briefly turn my attention to emission of Very High Energy (VHE) $\gamma$-rays from PKS 1510-089. The detection of VHE $\gamma$-ray photons from FSRQs is difficult to accommodate in a pure BLR-origin model for the location of the $\gamma$-ray emission region. VHE $\gamma$-ray emission from PKS 1510-089 was originally detected by the HESS telescope array in 2010 and has been recently confirmed by the MAGIC telescope array (\cite{hess}, \cite{magic}). A \textsc{gtlike} likelihood analysis of all $>50$ GeV photons from PKS 1510-089 detected by \textit{Fermi}-LAT, utilizing the same point source and diffuse model as that of \textsection 2, finds PKS 1510-089 to be a source of VHE $\gamma$-rays at the $\sim4.6\sigma$ confidence level. Furthermore, a \textsc{gtlike} likelihood analysis finds PKS 1510-089 to be a source of $\geq 20$ GeV $\gamma$-rays at the $\sim13.5\sigma$ confidence level.

A closer inspection of the individual photon events reveals that the highest energy photon detected from PKS 1510-089 has an energy of 70 GeV, with a total of 11 events above 20 GeV within 0.1 degrees of PKS 1510-089\footnote{0.1 degrees is larger than the 95\% containment angle for photons $\geq 20$ GeV.}. \cite{kat} and \cite{mengc} found that it is the $\geq $GeV $\gamma$-ray flux and spectral shape that is important when triggering ground-based VHE $\gamma$-ray observations, with a higher $\geq $GeV flux or, harder $\gamma$-ray spectrum, more likely to be associated with the emission of VHE $\gamma$-ray photons. To investigate if this applies to PKS 1510-089, the arrival times of the $\geq 20$ GeV photons was compared to the hardness ratio of the light curves of Figure \ref{lc2}. The hardness ratio was defined as the $1-300$ GeV flux divided by the $0.1-1$ GeV flux. The light curve of the hardness ratio, binned in daily intervals, along with the arrival times of the $\geq 20$ GeV photons, can be seen in Figure \ref{hardphotons}. 

There is no evidence in Figure \ref{hardphotons} of a correlation between the arrival of $\geq 20$ GeV photons and a spectral hardening as shown by a higher hardness ratio. Furthermore, there does not appear to be any obvious temporal clustering of $\geq 20$ GeV photons and the hardness ratio only exceeds 0.2 on one occasion. Interestingly, there are no $\geq 20$ GeV photons around the time where there is a spike in the hardness ratio of $\sim0.6$.

The MAGIC detection of PKS 1510-089 resulted from 10 hours of observations taken during the period 2012 February 3 to 2012 February 20 ($55960 <$ MJD $<55980$), either side of a full moon period (\cite{magic}). From Figure \ref{hardphotons} we see that there is no obvious increase in the hardness ratio during this period, and no $\geq 20$ GeV photons were detected either. Likewise, during the HESS observations, March 2010, only one $\geq20$ GeV photon was detected, and the hardness ratio exhibits a slight decreasing trend. As such, it would appear that for PKS 1510-089, there is no obvious $0.1 - 300$ GeV flux trend that leads to the emission of VHE $\gamma$-rays. This can be interpreted as further evidence of multiple emission regions along the jet being simultaneously active, since such a scenario would lead to complex variability patterns, resulting in an apparent random variation of the hardness ratio, that would mask an underlying pattern of VHE emission occuring when the $\geq $GeV $\gamma$-ray flux increases.  

\section{Conclusions}

An in-depth study of the $0.1-300$ GeV $\gamma$-ray properties of PKS 1510-089 utilising \textit{Fermi}-LAT observations from August 2008 to May 2012 has been reported. During this period, PKS 1510-089 has persistently been one of the brightest and most variable AGN observed by the LAT detector. Several large $\gamma$-ray flares were observed, where the daily $0.1-300$ GeV $\gamma$-ray flux exceeded $10^{-5}$ photons cm$^{-2}$ s$^{-1}$. The photon statistics associated with these large flares allowed the daily-binned light curve to be re-analysed in both 6 and 3 hour intervals and still satisfy a $TS \geq 10$ criteria for the vast majority of bins. The 3 hour binned light curve revealed the presence of doubling/halving timescales as small as $\tau_{rise}=1.30 \pm 0.12$ hours and $\tau_{decay}=1.21 \pm 0.15$ hours. 

The $10^{-5}$ photons cm$^{-2}$ s$^{-1}$ flare periods were studied in more detail in an attempt to find evidence for the location of PKS 1510-089's $\gamma$-ray emission region. In particular, two approaches were used: (i) searching for an energy dependence to the cooling timescales, and (ii) searching for evidence of a spectral cut-off. 

The flare 1 event ($55850 <$ MJD $< 55858$) possessed a spectral cut-off at its peak flux. However, the uncertainty in the DCF did not allow any clear conclusions on whether an energy dependency in the variability timescales is present. While there is an apparent peak in the DCF for flare 1 at a time-lag of 0, possibly hinting at a BLR origin for the flare, the large uncertainty in the DCF at negative time-lag values does not allow for any strong conclusions to be drawn from the DCF with regards to the origin of the $\gamma$-ray emission.

The flare 2 event ($55866 <$ MJD $< 55876$) showed no evidence for a spectral cut-off, possibly hinting at a MT origin of the $\gamma$-ray emission. If this is so, one would expect to see an energy dependency to the cooling timescale. While no evidence for this could be seen in the DCF of the $0.1-1$ GeV and $1-300$ GeV flux levels, this may to be due to the fact that the DCF is dominated by an apparent energy dependence in the flare rise time. 

These seemingly contradictory origins for the observed $\gamma$-ray flares can be reconciled if there are multiple, simultaneously active $\gamma$-ray emission zones along PKS 1510-089's relativistic jet, capable of emitting $0.1-300$ GeV $\gamma$-rays. Since the inner regions of FSRQs are opaque to $\gamma$-rays above $10-20$ GeV in energy, this multi-zone model can explain the existence of VHE $\gamma$-rays for PKS 1510-089, with emission regions in the MT producing the observed VHE flux, while those inside the BLR contributing to the $0.1-10$ GeV flux. While this study has unvealed evidence that both the BLR and MT location models are correct, further studies are needed to see if this property is unique to PKS 1510-089 or a characteristic of FSRQs in general.

\section*{Acknowledgments}

I thank SEB for her helpful discussions and insights which have been invaluable to this paper. I also thank the referee for her/his comments and suggestions that improved the quality and clarity of this paper. This work is supported by the Marsden Fund Council from New Zealand Government funding, administered by the Royal Society of New Zealand. This work has made use of public \textit{Fermi} data obtained from the High Energy Astrophysics Science Archive Research Center (HEASARC), provided by NASA’s Goddard Space Flight Center. This work has also made use of the NASA/IPAC Extragalactic Database (NED), which is operated by the Jet Propulsion Laboratory, Caltech, under contact with the National Aeronautics and Space Administration.

\end{document}